\documentclass[twocolumn,preprintnumbers,amsmath,amssymbm,prl]{revtex4}
\usepackage{epsfig}
\usepackage{graphicx}

\begin{document}
\title{From Thermodynamics to the Bound on Viscosity}
\author{Shahar Hod}
\address{The Ruppin Academic Center, Emeq Hefer 40250, Israel}
\address{ }
\address{The Hadassah Institute, Jerusalem 91010, Israel}
\date{\today}

\begin{abstract}
\ \ \ We show that the generalized second law of thermodynamics may
shed much light on the mysterious Kovtun-Son-Starinets (KSS) bound
on the ratio of viscosity to entropy density. In particular, we
obtain the lower bound $\eta/s +O(\eta^3/s^3)\geq 1/4\pi$.
Furthermore, for conformal field theories we obtain a new
fundamental bound on the value of the relaxation coefficient
$\tau_{\pi}$ of causal hydrodynamics, which has been the focus of
much recent attention: $(\tau_{\pi}T)^2\geq
{{(\sqrt{3}-1)}/{2\pi^2}}$.
\end{abstract}
\bigskip
\maketitle


The anti-de Sitter/conformal field theory (AdS/CFT) correspondence
\cite{Mald,Gub,Witt1,Brig1} has yielded remarkable insights into the
dynamics of strongly coupled gauge theories. According to this
duality, asymptotically AdS background spacetimes with event
horizons are interpreted as thermal states in dual field theories.
This implies that small perturbations of a black hole or a black
brane background correspond to small deviations from thermodynamic
equilibrium in a dual field theory. One robust prediction of the
AdS/CFT duality is a universally small ratio of the shear viscosity
to the entropy density \cite{Poli,Kov1,Buc,Kov2},
\begin{equation}\label{Eq1}
{\eta \over s}={1 \over{4\pi}}\  ,
\end{equation}
for all gauge theories with an Einstein gravity dual in the limit of
large 't Hooft coupling. (We use natural units for which
$G=c=\hbar=k_B=1$.)

It was suggested \cite{Kov2} that (\ref{Eq1}) acts as a universal
lower bound [the celebrated Kovtun-Starinets-Son (KSS) bound] on the
ratio of the shear viscosity to the entropy density of general,
possibly nonrelativistic, fluids. Currently this bound is considered
a {\it conjecture} well supported for a certain class of field
theories-- see the detailed discussions in \cite{Cher,BekFoux} and
the references therein. So far, all known materials satisfy the
bound for the range of temperatures and pressures examined in the
laboratory. The system coming closest to the bound is the
quark-gluon plasma created at the BNL Relativistic Heavy Ion
Collider (RHIC) \cite{Tean,Adar,Roma1,Song}. [In fact, it was the
challenge presented by the quark-gluon plasma which motivated the
activity leading to the formulation of the KSS bound (1).] Other
systems coming close to the bound include superfluid helium and
trapped ${^6}$Li at strong coupling \cite{Scha,Rup}. For other
related works, see \cite{SonStar,Mat,Dob,Den,Hern} and references
therein.

It is important to note that recent work \cite{Brig2} has shown
that, for a class of conformal field theories with Gauss-Bonnet
gravity dual, the shear viscosity to entropy density ratio,
$\eta/s$, could violate the conjectured KSS bound. In particular,
for $(3+1)$-dimensional CFT duals of $(4+1)$-dimensional
Gauss-Bonnet gravity, the ratio $\eta/s$ is given by
\begin{equation}\label{Eq2}
{\eta \over s} = {1 \over{4\pi}}(1-4\lambda_{GB})\  ,
\end{equation}
where $\lambda_{GB}$ is the Gauss-Bonnet coupling parameter
\cite{Zwie}. It was later shown that consistency of the theory
requires $\lambda_{GB}\leq {9 \over 100}$ \cite{Brig1}. This still
leaves rooms for a violation of the KSS bound (\ref{Eq1}). This
observation suggests that, if there is indeed a {\it universal}
lower bound on the ratio $\eta/s$, then it is likely to be a bit
more liberal than the originally conjectured KSS bound. This is
exactly the kind of result we shall find below. It should be noted,
however, that there is no known quantum field theory whose
hydrodynamic regime coincides with particularly chosen gravitational
lagrangian. So it might also be the case that the KSS bound is
robust and the Gauss-Bonnet lagrangian does not capture the
hydrodynamic limit of a field theory.

Where does the KSS bound (or any other refined bound on the ratio
$\eta/s$) come from? It is not clear how to obtain such a bound
directly from microscopic physics \cite{BekFoux}. Inspection of the
Green-Kubo formula \cite{Land} which relates the viscosity of a
fluid to its fluctuations shows no apparent connection of the
viscosity and the entropy density. Such microscopic consideration
affords no special status to the ratio $\eta/s$ \cite{BekFoux}.

Where should we look for the physical mechanism which bounds the
ratio $\eta/s$ of viscosity to entropy density? It is well known
that the viscosity coefficient $\eta$ characterizes the intrinsic
ability of a perturbed fluid to relax towards equilibrium
\cite{Dani} [see Eqs. (\ref{Eq3}) and (\ref{Eq10}) below]. The
response of a medium to mechanical excitations is characterized by
two types of normal modes, corresponding to whether the momentum
density fluctuations are transverse or longitudinal to the fluid
flow. Transverse fluctuations lead to the shear mode, whereas
longitudinal momentum fluctuations lead to the sound mode. (There is
also the diffuse mode in the presence of a conserved current.) These
perturbation modes are characterized by distinct dispersion
relations which describe the poles positions of the corresponding
retarded Green functions \cite{SonStar}.

Let us first examine the behavior of the shear mode for fluids with
zero chemical potential. The Euler identity reads $\epsilon+P=Ts$,
where $\epsilon$ is the energy density, $P$ is the pressure, $T$ is
the temperature, and $s$ is the entropy density of the fluid. The
dispersion relation for a shear wave with frequency $\omega$ and
wave vector $k\equiv 2\pi/\lambda$ is given by \cite{Baier,Nats}:
\begin{equation}\label{Eq3}
\omega(k)_{\text{shear}}=-i{{\eta}\over{Ts}}k^2+O\Big({{\eta^3k^4}\over{s^3T^3}}\Big)\
,
\end{equation}
where $\eta$ is the shear viscosity coefficient of the standard
first-order hydrodynamics. The correction term becomes small in the
$\eta/s\ll 1$ limit, the case of most interest here.

The imaginary part of the dispersion relation entails a damping of
the perturbation mode. Its magnitude therefore quantifies the
intrinsic ability of a fluid to dissipate perturbations and to
approach thermal equilibrium.

It is important to realize that hydrodynamics is actually an
effective theory. In the most common applications of hydrodynamics
the underlying microscopic theory is a kinetic theory. In this case
the microscopic scale which limits the validity of the effective
hydrodynamic description is the mean free path $l_{\text{mfp}}$
\cite{Baier}. More generally, the underlying microscopic theory is a
quantum field theory, which might not necessarily admit a kinetic
description. In these cases, the role of the parameter
$l_{\text{mfp}}$ is played by some typical microscopic scale like
the inverse temperature: $\l_{\text{mfp}}\sim T^{-1}$. One therefore
expects to find a breakdown of the effective hydrodynamic
description at spatial and temporal scales of the order of
\cite{Baier}
\begin{equation}\label{Eq4b}
l\sim \tau\sim T^{-1}\  .
\end{equation}
Below we shall make this statement more accurate. It should be noted
that the relation (\ref{Eq4b}) may be modulated by a function of the
dimensionless parameters of the theory (if any).

At this point, it is worth emphasizing that the conjectured KSS
bound is based on holographic calculations of the shear viscosity
for strongly coupled quantum field theories with gravity duals.
These holographic arguments serve to connect quantum field theory
with gravity. This fact indicates that a derivation of a KSS-like
bound may require use of the still nonexistent theory of quantum
gravity \cite{BekFoux}. This may seem as bad news for our
aspirations to prove (a refined version of) the KSS bound. But one
need not loose heart-- there is general agreement that black hole
entropy reflects some aspect of the elusive theory of quantum
gravity \cite{BekFoux}.

The realization that a black hole is endowed with well-defined
entropy $S_{BH}=A/4$, where $A$ is the surface area of the black
hole \cite{Bek1,Haw}, has lead to the formulation of the generalized
second law (GSL) of thermodynamics. The GSL is a unique law of
physics that bridges thermodynamics and gravity
\cite{Bek1,Haw,BekFoux}. It asserts that in any interaction of a
black hole with an ordinary matter, the sum of the entropies
(matter+hole) never decreases. One of the most remarkable
predictions of the GSL is the existence of a universal entropy bound
\cite{Bek4,Bek5}. According to this universal bound, the entropy
contained in a given volume should be bounded from above:
\begin{equation}\label{Eq5}
S\leq 2\pi RE\  ,
\end{equation}
where $R$ is the effective radius of the system and $E$ is its total
energy.

Furthermore, the generalized second law allows one to derive in a
simple way two important new quantum bounds:
\begin{itemize}
\item{The universal relaxation bound \cite{Hod1,Hod2,Gruz}. This bound asserts that the
relaxation time of a perturbed thermodynamic system is bounded from
below by
\begin{equation}\label{Eq6}
\tau\geq 1/\pi T\  ,
\end{equation}
where $T$ is the temperature of the system. This bound can be
regarded as a quantitative formulation of the third law of
thermodynamics. One can also write this bound as $\Im {\varpi}\leq
1/2$, where ${\varpi}\equiv \omega/2\pi T$. The connection between
the universal relaxation bound (\ref{Eq6}) and the Bekenstein
entropy bound (\ref{Eq5}) is established in Ref. \cite{Hod1}.}
\item{A closely related conclusion is that thermodynamics can not be
defined on arbitrarily small length scales. The minimal length scale
(radius) $\ell$ for which a consistent thermodynamic description is
available is given by $\ell_{min}=1/2\pi T$, see Refs.
\cite{Pes1,Pes2,BekFoux}.}
\end{itemize}

The longest wavelength which can fit into a space region of
effective radius $\ell$ is $\lambda_{max}=2\pi\ell$. Thus, the GSL
predicts that an effective hydrodynamic description is limited to
perturbation modes with wavelengths larger than
$2\pi\ell_{min}=T^{-1}$. This limit agrees with the one found from
the heuristic argument presented above \cite{Baier}. The breakdown
of the effective hydrodynamic description for perturbation modes
with wavenumbers $k$ larger than $2\pi T$ should manifest itself in
the hydrodynamic dispersion relation (\ref{Eq3}) [see also Eq.
(\ref{Eq8}) below]. This breakdown may reveal itself in two distinct
ways:
\begin{itemize}
\item Short relaxation times which violate the universal relaxation bound
(\ref{Eq6}), or
\item Superluminal sound propagation which violates causality
(this is not relevant for the shear mode).
\end{itemize}

A lower bound on the ratio $\eta/s$ can be inferred by substituting
${\rm q}\equiv k/2\pi T=1$ in the shear dispersion relation
(\ref{Eq3}) and requiring that $\Im\varpi\geq 1/2$ for this limiting
value of the wavenumber. As discussed above, the GSL predicts that
the effective hydrodynamic description breaks down for short
wavelength perturbations with ${\rm q}>1$. This should be reflected
in the hydrodynamic shear dispersion relation in the form of a
violation of the universal relaxation bound (\ref{Eq6}). Explicitly,
these wavenumbers should be characterized by $\Im\varpi>1/2$. This
physical condition leads to the simple bound
\begin{equation}\label{Eq7}
{\eta \over s}+O\Big({{\eta^3}\over{s^3}}\Big)\geq {1 \over{4\pi}}\
.
\end{equation}

Let us now examine the sound perturbation mode. The sound dispersion
relation for conformal field theories is given by
\cite{Baier,Nats,Notebul}:
\begin{equation}\label{Eq8}
\Re\omega(k)_{\text{sound}}\simeq\pm v_s k\pm
{\Gamma\over{v_s}}\Big(v^2_s\tau_{\pi}-{\Gamma\over 2}\Big)k^3\  ,
\end{equation}
where $v_s=\sqrt{dP/d\epsilon}=1/\sqrt{d}$ ($d$ is the number of
spatial dimensions), $\Gamma={{d-1}\over{d}}{\eta\over{Ts}}$, and
$\tau_{\pi}$ is a relaxation coefficient whose origin is in the
second-order causal hydrodynamics (see details below). The imaginary
part of the sound dispersion relation is given by
\begin{equation}\label{Eq10}
\Im\omega(k)_{\text{sound}}=-i{{d-1}\over d}{{\eta}\over{Ts}}
k^2+\cdots\ .
\end{equation}

A frequently used formalism for second-order hydrodynamics is the
`M\"uller-Israel-Stewart' formalism \cite{Mul,Isr,IsrSte}. This
extension of the first-order hydrodynamics attempts to repair
problems that the first order theory has with causality, and
necessarily introduces a set of new transport coefficients like
$\tau_{\pi}$. This coefficient has the dimension of time, and it is
often referred to as a relaxation time (although that is somewhat of
a misnomer \cite{Kapu}). This new transport coefficient has been the
focus of much recent attention. In particular, several groups have
calculated this coefficient for various strongly coupled field
theories \cite{Baier,Nats,Heller,Hub,Beni}.

The expected breakdown of the effective hydrodynamic description for
sound mode perturbations with short wavelengths ${\rm q}>1$ can
manifest itself in two distinct ways: (1) A superluminal sound
propagation with $v_g=d\Re\varpi/d{\rm q}>1$, or (2) Short
relaxation times (characterized by $\Im\varpi>1/2$) which violate
both the universal relaxation bound (\ref{Eq6}) and the GSL. Either
one of these two options is by itself sufficient to infer a
breakdown of the effective hydrodynamic description. Taking
cognizance of the dispersion relations (\ref{Eq8}) and (\ref{Eq10}),
and requiring that $v_g\geq 1$ or $\Im\varpi\geq 1/2$ for the
limiting wavenumber ${\rm q}=1$, one obtains the lower bounds
\begin{equation}\label{Eq11} {\eta \over
s}\Big(\tau_{\pi}T-{\eta\over s}\Big)\geq {{{\sqrt
3}-1}\over{8\pi^2}}\ \ \ {\text or}\ \ \ {\eta \over
s}+O\Big({{\eta^3}\over{s^3}}\Big)\geq {3 \over{8\pi}}\  ,
\end{equation}
for the physically interesting case of field theories in three
spatial dimensions.

It is instructive to check the validity of this new bound
(\ref{Eq11}) against known results. For example, the canonical model
of strongly coupled finite temperature ${\cal N}=4$ supersymmetric
$SU(N_c)$ Yang Mills theory in the limit of large $N_c$ is
characterized by the well-known ratio $\eta/s=1/4\pi$. Most
recently, several groups have calculated the relaxation coefficient
$\tau_{\pi}$ of the second-order causal hydrodynamics and found
$\tau_{\pi}T=(2-\ln2)/2\pi$ for this model
\cite{Baier,Nats,Heller,Hub,Beni}. Substituting these two values
into the l.h.s of (\ref{Eq11}), one may directly confirm the
validity of the inequality. In fact, this canonical model is
remarkably close ($\sim 4\%$) of saturating the bound (\ref{Eq11}).

The new bound (\ref{Eq11}) combines the viscosity coefficient $\eta$
of first-order hydrodynamics with the relaxation coefficient
$\tau_{\pi}$ of second-order hydrodynamics. From this bound one may
also infer a concrete lower bound on the value of the relaxation
coefficient $\tau_{\pi}$ of the second-order causal hydrodynamics:
\begin{equation}\label{Eq13}
(\tau_{\pi}T)^2\geq {{\sqrt{3}-1}\over{2\pi^2}}\  .
\end{equation}
This inequality should be satisfied by theories characterized by
$\eta/s<3/8\pi$.

In summary, we have given support to the idea that a lower bound on
the viscosity to entropy ratio $\eta/s$ may possibly be inferred
from the generalized second law of thermodynamics. The bound
(\ref{Eq7}) may be a bit more liberal than the originally
conjectured KSS bound (\ref{Eq1}). Being a direct consequence of the
generalized second law of thermodynamics, this bound is expected to
be of general validity.


\bigskip
\noindent
{\bf ACKNOWLEDGMENTS}
\bigskip

This research is supported by the Meltzer Science Foundation. I
thank R. Baier, P. Romatschke, D. T. Son, A. O. Starinets, M. A.
Stephanov, and A. Pesci for helpful correspondence. I also thank
Yael Oren for stimulated discussions.



\begin{thebibliography}{99}

\bibitem{Mald} J. M. Maldacena, Adv. Theor. Math. Phys. {\bf 2}, 231
(1998); [Int. J. Theor. Phys. {\bf 38}, 1113 (1999)].

\bibitem{Gub} S. S. Gubser, I. R. Klebanov, and A.  M. Polyakov,
Phys. Lett. B {\bf 428}, 105 (1998).

\bibitem{Witt1} E. Witten, Adv. Theor. Math. Phys. {\bf 2}, 253
(1998); E. Witten, Adv. Theor. Math. Phys. {\bf 2}, 505 (1998).

\bibitem{Brig1} M. Brigante, H. Liu, R. C. Myers, S. Shenker, and S.
Yaida, Phys. Rev. Lett. {\bf 100}, 191601 (2008).

\bibitem{Poli} G. Policastro, D. T. Son, and A. O. Starinets, Phys.
Rev. Lett. {\bf 87}, 081601 (2001).

\bibitem{Kov1} P. Kovtun, D. T. Son, and A. O. Starinets, J. High
Energy Phys. 10, (2003) 064.

\bibitem{Buc} A. Buchel and J. T. Liu, Phys. Rev. Lett. {\bf 93}, 090602 (2004).

\bibitem{Kov2} P. Kovtun, D. T. Son, and A. O. Starinets, Phys. Rev. Lett. {\bf
94}, 111601 (2005).

\bibitem{Cher} A. Cherman, T. D. Cohen, and P. M. Hohler, J. High
Energy Phys. 0802:026 (2008).

\bibitem{BekFoux} I. Fouxon, G. Betschart, and J. D. Bekenstein,
Phys. Rev. D {\bf 77}, 024016 (2008).

\bibitem{Tean} D. Teaney, Phys. Rev. C {\bf 68}, 034913 (2003);
K. Dusling and D. Teaney, Phys. Rev. C {\bf 77},
034905 (2008).

\bibitem{Adar} A. Adare {\it et al.} (PHENIX Collaboration), Phys.
Rev. Lett. {\bf 98}, 172301 (2007)

\bibitem{Roma1} P. Romatschke and U. Romatschke, Phys. Rev. Lett.
{\bf 99}, 172301 (2007); P. Romatschke, arXiv:0710.0016.

\bibitem{Song} H. Song and U. W. Heinz, Phys. Lett. B {\bf 658}, 279
(2008).

\bibitem{Scha} T. Schafer, Phys. Rev. A {\bf 76}, 063618 (2007).

\bibitem{Rup} G. Rupak and T. Schafer, Phys. Rev. A {\bf 76}, 053607
(2007).

\bibitem{SonStar} D. T. Son and A. O. Starinets, Ann. Rev. Nucl. Part.
Sci. {\bf 57}, 95 (2007).

\bibitem{Mat} D. Mateos, R. C. Myers, and R. M. Thomson, Phys. Rev.
Lett. {\bf 98}, 101601 (2007).

\bibitem{Dob} A. Dobado, F. J. Llanes-Estrada, and J. M. T.
Rinc\'on, arXiv:0804.2601.

\bibitem{Den} G. S. Denicol, T. Kodama, T. Koide, and Ph. Mota,
arXiv:0807.3120.

\bibitem{Hern} C. P. Herzog, N. Rangamani, S. F. Ross, J. High Energy Phys. 0811:080
(2008).

\bibitem{Brig2} M. Brigante, H. Liu, R. C. Myers, S. Shenker, and S.
Yaida, Phys. Rev. D {\bf 77}, 126006 (2008).

\bibitem{Zwie} B. Zwiebach, Phys. Lett. B {\bf 156}, 315 (1985).

\bibitem{Land} L. D. Landau and E. M. Lifshitz, {\it Statistical
Physics} (Butterworth-Heinemann, Oxford, 2000), Vol. II.

\bibitem{Dani} P. Danielewicz and M. Gyulassy, Phys. Rev. D {\bf
31}, 53 (1985).

\bibitem{Baier} R. Baier, P. Romatschke, D. T. Son, A. O. Starinets, and M. A.
Stephanov, J. High Energy Phys. 0804:100 (2008).

\bibitem{Nats} M. Natsuume and T. Okamura, Phys. Rev. D {\bf 77},
066014 (2008).

\bibitem{Bek1} J. D. Bekenstein, Lett. Nuovo Cimento Soc. Ital. Fis.
{\bf 4}, 737 (1972); J. D. Bekenstein, Phys. Rev. D {\bf 7}, 2333
(1973); J. D. Bekenstein, Phys. Rev. D {\bf 9}, 3292 (1974);

\bibitem{Haw} S. W. Hawking, Commun. Math. Phys. {\bf 43}, 199
(1975).

\bibitem{Bek4} J. D. Bekenstein, Phys. Rev. D {\bf 23}, 287 (1981).

\bibitem{Bek5} J. D. Bekenstein, Found. Phys. {\bf 35}, 1805 (2005).

\bibitem{Hod1} S. Hod, Phys. Rev. D {\bf 75}, 064013 (2007).

\bibitem{Hod2} S. Hod, Class. Quantum Grav. {\bf 24}, 4235 (2007).

\bibitem{Gruz} A. Gruzinov, arXiv:0705.1725.

\bibitem{Pes1} A. Pesci, Class. Quantum Grav. {\bf 24}, 6219 (2007).

\bibitem{Pes2} A. Pesci, arXiv:0807.0300.

\bibitem{Amado} I. Amado, C. Hoyos, K. Landsteiner, and S. Montero,
arXiv:0805.2570.

\bibitem{Notebul} For CFTs it is possible to use conformal Ward
identities to show that the bulk viscosity vanishes: $\zeta=0$
\cite{SonStar}.

\bibitem{Mul} I. M\"uller, Z. Phys. {\bf 198}, 329 (1967).

\bibitem{Isr} W. Israel. Ann. Phys. (NY) {\bf 100}, 310 (1976).

\bibitem{IsrSte} W. Israel and J. M. Stewart, Ann. Phys. (NY) {\bf 118}, 341
(1979).

\bibitem{Kapu} J. I. Kapusta and T. Springer, arXiv:0806.4175.

\bibitem{Heller} M. P. Heller and R. A. Janik, Phys. Rev. D
{\bf 76}, 025027 (2007).

\bibitem{Hub} S. Bhattacharyya, V. E. Hubeny, S. Minwalla, and M.
Rangamani, J. High Energy Phys. 0802:045 (2008).

\bibitem{Beni} P. Benincasa, A. Buchel, M. P. Heller, R. A. Janik,
Phys. Rev. D {\bf 77}, 046006 (2008).

\end{thebibliography}
\end{document}